\documentclass[twocolumn,pra,showpacs,superscriptaddress,floatfix]{revtex4-1}

\usepackage{graphicx}
\usepackage[latin1]{inputenc}
\usepackage{graphicx,epstopdf}
\usepackage{dcolumn}
\usepackage{amsmath}
\usepackage[hypertex]{hyperref}
\usepackage{amsfonts}
\usepackage{amsthm}
\usepackage{amssymb}
\usepackage{mathrsfs}
\PassOptionsToPackage{caption=false}{subfig}
\usepackage{subfig}
\usepackage{color}
\usepackage[normalem]{ulem}

\begin{document}

\newcommand{\red}[1]{\textcolor{red}{#1}}
\newcommand{\blue}[1]{\textcolor{blue}{#1}}

\newcommand{\inprod}[2]{\braket{#1}{#2}}
\newcommand{\Tr}{\mathrm{Tr}}
\newcommand{\vp}{\vec{p}}
\newcommand{\Or}{\mathcal{O}}
\newcommand{\so}[1]{{\ignore{#1}}}

\title{Trace-distance correlations for X states and emergence of the pointer basis in Markovian and non-Markovian regimes} 
\author{Paola C. Obando}
\email{paolaconcha@if.uff.br}
\affiliation{Instituto de F\'{\i}sica, Universidade Federal Fluminense, Avenida General Milton Tavares de Souza s/n, Gragoat\'a, 24210-346,Niter\'oi, RJ, Brazil}

\author{Fagner M. Paula}
\email{fagnerm@utfpr.edu.br}
\affiliation{Universidade Tecnol\'ogica Federal do Paran\'a - Rua Cristo rei 19, Vila Becker, 85902-490, Toledo, PR, Brazil}

\author{Marcelo S. Sarandy}
\email{msarandy@if.uff.br}
\affiliation{Instituto de F\'{\i}sica, Universidade Federal Fluminense, Avenida General Milton Tavares de Souza s/n, Gragoat\'a, 24210-346,Niter\'oi, RJ, Brazil}

\date{\today}

\begin{abstract}

We provide analytical expressions for classical and total trace-norm (Schatten 1-norm) geometric correlations in the case of two-qubit X states. As an application, we consider 
the open-system dynamical behavior of such correlations under phase and generalized amplitude damping evolutions. Then, we show that geometric classical correlations can 
characterize the emergence of the pointer basis of an apparatus subject to decoherence in either Markovian or non-Markovian regimes. In particular, as a non-Markovian effect, 
we obtain a time delay for the information to be retrieved from the apparatus by a classical observer. Moreover, we show that the set of initial X states exhibiting sudden transitions 
in the geometric classical correlation has nonzero measure. 

\end{abstract}

\pacs{03.65.Ud, 03.67.Mn, 75.10.Jm}

\maketitle

\section{Introduction}

Correlations are typically behind information-based interpretations of physical phenomena~\cite{Modi,Celeri,Sarandy:12}. In a quantum scenario, 
they appear as key signatures, with operational roles e.g.~in quantum metrology \cite{modix,lqu,blind}, entanglement activation \cite{streltsov,Piani:11,sciarrino}, 
and information encoding and distribution \cite{gu,sharing}. In a geometric approach, they can be defined  through a number of distinct formulations, which are based 
on the relative entropy~\cite{Modi:10}, Hilbert-Schmidt norm~\cite{Dakic:10,Bellomo:12}, trace norm~\cite{Paula,Nakano:12}, or Bures norm~\cite{Spehner:13,Bromley}. 
All of these distinct versions can be generally described by a unified framework in terms of a distance (or pseudo distance) function. In particular, it has been shown that 
the trace norm, which corresponds to the Schatten $1$-norm, provides a suitable direction for the investigation of quantum, classical, and total correlations,  since it is 
the only $p$-norm able to satisfy  reasonable axioms expected to hold for information-based correlation functions. Moreover, for the simple case of mixed two-qubit 
systems in Bell-diagonal states, analytical expressions have been found for quantum, classical, and total correlations~\cite{EPL103,Aaronson:13,EPL108}. However, for 
the more general case of two-qubit X states, only the quantum contribution for the geometric correlation has been analytically derived~\cite{ciccarello}. Here, our aim is 
to close this gap, providing closed analytical expressions for the classical and total correlations of arbitrary two-qubit X states. Remarkably, they are shown to be as simple to be 
computed as in the case of Bell-diagonal states. 

The analytical expressions for the classical correlation of X states can be applied as a powerful resource to characterize the open-system 
dynamics in rather general environments. In this direction, we consider a system-apparatus set ${\cal AS}$ under the effect of X state preserving channels, with decoherence driving  
the quantum apparatus ${\cal A}$ to collapse into a possible set of classical states known as the pointer basis~\cite{zurek}. We are then able to show that the geometric classical 
correlation decays to a constant value at finite time $\tau_E$ for decohering processes admitting a pointer basis. This is exploited in a general scenario of X states, for either 
Markovian and non-Markovian evolutions. In particular, we show a delay  in the emergence time $\tau_E$ in the non-Markovian regime. 

\section{Geometric classical and total correlations: analytical expressions}

In the general approach introduced in Refs.~\cite{Modi,brodutch-m,EPL108}, measures of quantum, classical, 
and total correlations of an $n$-partite system in a state $\rho$ are respectively defined by 
\begin{equation}
Q(\rho)=K\left[\rho,M_{-}(\rho)\right],
\end{equation}
\begin{equation}
C(\rho)=K\left[M_{+}(\rho),M_{+}(\pi_{\rho})\right],
\end{equation}
\begin{equation}
T(\rho)=K\left[\rho,\pi_{\rho}\right],
\end{equation}  
where $K\left[\rho,\tau\right]$ denotes a real and positive function that vanishes for $\rho=\tau$, $M_{-}(\rho)$ is a classical state obtained through a non-selective 
measurement $\{ M_{-}^{(i)}\}$ that minimizes $Q$, $M_{+}(\rho)$ is a classical state obtained through a non-selective measurement $\{ M_{+}^{(i)}\}$ that maximizes $C$,  
and $\pi_{\rho}=\rho_1\otimes...\otimes\rho_n = \text{tr}_{\bar{1}}\rho\otimes...\otimes\text{tr}_{\bar{n}}\rho$ represents the product of the local marginals  of $\rho$. 
In order to avoid ambiguities in the correlation measures for $Q$ and $C$, we take $\{ M_{-}^{(i)}\}$ and $\{M_{+}^{(i)}\}$ as independent measurement sets~\cite{EPL108}. 
Let us consider correlations based on the trace norm (Schatten 1-norm) and projective measurements operating over one qubit of a two-qubit system, i.e., 
$K\left[\rho,\tau\right]=\left\Vert \rho-\tau\right\Vert_{1}=\mathrm{tr}|\rho-\tau|$ and $M_{\pm}(\rho)=\Pi_{\pm}^{(1)}(\rho)$, such that
\begin{equation}
Q_G(\rho)=\mathrm{tr}\left|\rho-\Pi_{-}^{(1)}(\rho)\right|,
\end{equation}
\begin{equation}\label{cg-0}
C_G(\rho)=\mathrm{tr}\left|\Pi_{+}^{(1)}(\rho)-\Pi_{+}^{(1)}(\pi_{\rho})\right|,
\end{equation}
\begin{equation}\label{tg}
T_G(\rho)= \mathrm{tr}\left|\rho-\pi_{\rho}\right|.
\end{equation}  
By adopting the trace norm, $Q_G$ is then the Schatten 1-norm geometric quantum discord, as introduced in Refs.~\cite{Paula,Nakano:12}. In particular, for two-qubit systems, 
the geometric quantum discord based on Schatten 1-norm is equivalent to the negativity of quantumness~\cite{Nakano:12} (also referred as the minimum entanglement 
potential~\cite{Chaves:11}), which is a measure of nonclassicality introduced in Ref.~\cite{Piani:11} and experimentally discussed in Ref.~\cite{Silva:12}. As a counterpart to 
$Q_G$, $C_G$ is the Schatten 1-norm classical correlation. Concerning $T_G$, it is a measure of total geometric correlation, which vanishes if the system is described by 
a product state. The trace norm satisfies reasonable criteria expected for correlation measures, although these criteria are still source of debate~\cite{EPL108,Maziero:15}.

We are interested in a two-qubit system as described by an X-shaped mixed state. Two-qubit X states describe rather general two-qubit systems. These states generalize the 
Bell-diagonal states, which are those whose density matrix is diagonal in the Bell basis. An example of a Bell-diagonal state (and therefore of an X state) is the 
Werner state~\cite{Werner:89}, which mixes a singlet (maximally entangled) state with the identity (fully classical) state. In condensed matter physics, X states provide the general form 
of reduced density operators of arbitrary quantum spin chains with $Z_2$ (parity) symmetry (for a review see, e.g., Ref.~\cite{Sarandy:12}). For example, both ground and thermal reduced 
two-spin states of the quantum Ising chain in a tranverse magnetic field are described by X states. The same holds for other spin chains, such Heisenberg and XXZ models. 
The density matrix of a two-qubit X state takes the form
\begin{equation}
\mathcal{\rho}_{X}=\left( 
\begin{array}{cccc}
\rho_{11} & 0 & 0 & \rho_{41}^{*} \\ 
0 & \rho_{22} & \rho_{32}^{*} & 0 \\ 
0 & \rho_{32} & \rho_{33} & 0 \\ 
\rho_{41} & 0 & 0 & \rho_{44}
\end{array}
\right),
\end{equation}
where computational basis $\{\left|00\right\rangle,\left|01\right\rangle,\left|10\right\rangle,\left|11\right\rangle\}$ is adopted. 
The normalization and the positive semidefiniteness of state require $\sum_{i=1}^{4}{\rho_{ii}}=1$,  $\rho_{11}\rho_{44}\geq|\rho_{41}|^{2}$, and $\rho_{22}\rho_{33}\geq|\rho_{32}|^{2}$. 
The diagonal elements are real, whereas the elements $\rho_{41}$ and $\rho_{32}$ are complex numbers in general. However, they can be brought into real numbers via local unitary 
transformations, which preserve the trace distance correlations~\cite{ciccarello}. By decomposing the X state in the Pauli basis, we obtain
\begin{equation}\label{eq:rhox}
\rho_{X}=\frac{1}{4}\left(\mathbb{I}\otimes \mathbb{I}+\sum_{i=1}^{3}c_{i}\sigma_{i}\otimes\sigma_{i}+c_{4}\mathbb{I}\otimes\sigma_{3}+c_{5}\sigma_{3}\otimes\mathbb{I}\right)
\end{equation}
where 
\begin{equation}\label{c1}
c_{1}=\text{tr}(\sigma_{1}\otimes\sigma_{1}\rho_{X})=2(\rho_{32}+\rho_{41}),
\end{equation}
\begin{equation}\label{c2}
c_{2}=\text{tr}(\sigma_{2}\otimes\sigma_{2}\rho_{X})=2(\rho_{32}-\rho_{41}),
\end{equation}
\begin{equation}\label{c3}
c_{3}=\text{tr}(\sigma_{3}\otimes\sigma_{3}\rho_{X})=1-2(\rho_{22}+\rho_{33}),
\end{equation}
\begin{equation}\label{c4}
c_{4}=\text{tr}(\mathbb{I}\otimes\sigma_{3}\rho_{X})=2(\rho_{11}+\rho_{33})-1,
\end{equation}
\begin{equation}\label{c5}
c_{5}=\text{tr}(\sigma_{3}\otimes\mathbb{I}\rho_{X})=2(\rho_{11}+\rho_{22})-1,
\end{equation}
with all these parameters assuming values in the interval $-1\leq c_i\leq 1$. If $c_{4}=c_{5}=0$, we obtain the Bell-diagonal state:
\begin{equation}
\rho_{X}=\rho_{B} \,\,\,\,\,\,\,\left(c_4=c_5=0\right).
\end{equation}
In terms of the parameters $\{c_i\}$, the Schatten 1-norm quantum correlation can be written as~\cite{ciccarello}
\begin{equation}
Q_{G}(\rho_{X})=\sqrt{\frac{ac-bd}{a-b+c-d}},
\end{equation}
where $a=\text{max}\{c_{3}^{2},d+c_{5}^{2}\}$, $b=\text{min}\{c,c_{3}^{2}\}$, $c=\text{max}\{c_{1}^{2},c_{2}^{2}\}$, and $d=\text{min}\{c_{1}^{2},c_{2}^{2}\}$.
Now, let us calculate the corresponding classical and total correlations. First, by computing the marginal density operators, we get 
$\rho_{1}=\text{tr}_{\bar{1}}\rho=\left(\mathbb{I}+c_5\sigma_{3}\right)/2$ and $\rho_{2}=\text{tr}_{\bar{2}}\rho=\left(\mathbb{I}+c_4\sigma_{3}\right)/2$. Then,  
the product state $\pi_{\rho_{X}}=\rho_1\otimes\rho_2$ reads
\begin{equation}
\pi_{\rho_{X}}= \frac{1}{4}\left(\mathbb{I}\otimes \mathbb{I}+c_{4}\mathbb{I}\otimes\sigma_{3}+c_{5}\sigma_{3}\otimes\mathbb{I}+c_{4}c_{5}\sigma_{3}\otimes\sigma_{3}\right).
\label{pi-X}
\end{equation}
From Eq.~(\ref{cg-0}), we observe that $\Pi_{+}^{(1)}(\rho)-\Pi_{+}^{(1)}(\pi_{\rho}) = \Pi_{+}^{(1)}(\rho-\pi_{\rho})$. 
Then, by using Eqs.~(\ref{eq:rhox})~and~(\ref{pi-X}), we 
can observe that the difference of X-states $\rho_X-\pi_{\rho_X}$ is mathematically equivalent to a difference between Bell-diagonal states. 
Indeed, we can rewrite the difference 
$\rho_X-\pi_{\rho_X}$ [also appearing in Eq.~(\ref{tg})] in terms of effective Bell-diagonal states $\tilde{\rho}_{B}$ and $\pi_{\tilde{\rho}_{B}}$, i.e., 
\begin{equation}
\rho_{X}-\pi_{\rho_{X}}=\tilde{\rho}_{B}-\pi_{\tilde{\rho}_{B}},
\end{equation}
where
\begin{equation}
\tilde{\rho}_{B}=\frac{1}{4}\left[\mathbb{I}\otimes \mathbb{I}+\sum_{i=1}^{3}\tilde{c}_{i}\sigma_{i}\otimes\sigma_{i}\right]
\end{equation}
and
\begin{equation}
\pi_{\tilde{\rho}_{B}}= \frac{1}{4}\left(\mathbb{I}\otimes \mathbb{I}\right),
\end{equation}
with
\begin{equation}\label{nuevasc}
\left(\tilde{c}_1,\tilde{c}_2,\tilde{c}_3\right)=\left(c_1,c_2,c_3-c_4c_5\right).
\end{equation}
In this case, we can directly apply the analytical expressions of $C_G$ and $T_G$ already obtained for the Bell-diagonal state \cite{EPL103,EPL108}. This procedure 
implies in the correlation measures for X states obtained in this work, which read
\begin{equation}\label{cg}
C_{G}(\rho_{X})=C_{G}(\tilde{\rho}_{B})=\tilde{c}_{+}
\end{equation}
and
\begin{equation}
T_{G}(\rho_{X})=T_{G}(\tilde{\rho}_{B})=\frac{1}{2}\left[\tilde{c}_{+}+\max\{\tilde{c}_{+},\tilde{c}_{0}+\tilde{c}_{-}\}\right],
\end{equation}
where $\tilde{c}_{-}=\text{min}\{|\tilde{c}_1|,|\tilde{c}_2|,|\tilde{c}_3|\}$, $\tilde{c}_{0}=\text{int}\{|\tilde{c}_1|,|\tilde{c}_2|,|\tilde{c}_3|\}$, and $\tilde{c}_{+}=\text{max}\{|\tilde{c}_1|,|\tilde{c}_2|,|\tilde{c}_3|\}$ 
represent the minimum, intermediate, and the maximum of the absolute values of the parameters $\tilde{c}_i$ ($i=1,2,3$), respectively.

\section{applications}

We illustrate the applicability of the geometric measure of classical correlations by considering the decohereing dynamics of the quantum systems. 
We will take the system as a two qubit state coupled independently with weak sources of noise ~\cite{nielsen} (either phase or generalized amplitude damping). 
This scenario appears in many situations, such as optical quantum systems~\cite{yu:93} and nuclear magnetic resonance (NMR) setups~\cite{PRL111}. 

\subsection{Markovian dynamics}

Let us consider a Markovian process as described by the operator-sum representation 
formalism~\cite{nielsen}. In this scenario, the evolution of a quantum state 
$\rho$ is governed by a trace-preserving quantum operation $\varepsilon(\rho)$, which is given by 
\begin{equation}
\varepsilon(\rho) = \sum_{i,j} \left(E_i^A\otimes E_j^B\right) \rho \left(E_i^A \otimes E_j^B\right)^\dagger, 
\label{epsilon_rho}
\end{equation}
where $\{E_k^s\}$ is the set of Kraus operators associated with a decohering process of a single qubit, 
with the trace-preserving condition reading $\sum_k E_k^{s\dagger} E_k^s = I$. We provide in Table~\ref{t1} 
the Kraus operators for phase damping (PD) and generalized amplitude damping (GAD), which are the channels 
considered in this work.

\begin{table}[hbt]
\begin{tabular}{|c|c|}
\hline
 & $\textrm{Kraus operators}$                                         \\ \hline \hline
 & \\ 
PD   & $E_0^s = \sqrt{1-p_s/2}\, I , E_1^s = \sqrt{p_s/2}\, \sigma_3$                        \\ \hline
 & \\ 
GAD   & 
$E_0^s=\sqrt{\lambda_s}\left( 
\begin{array}{cc}
1 & 0 \\ 
0 & \sqrt{1-p_s} \\ 
\end{array} \right) , 
E_2^s=\sqrt{1-\lambda_s}\left( 
\begin{array}{cc}
\sqrt{1-p_s} & 0 \\ 
0 & 1 \\ 
\end{array} \right)$  \\
& \\
 & $E_1^s=\sqrt{\lambda_s}\left( 
\begin{array}{cc}
0 & \sqrt{p_s} \\ 
0 & 0 \\ 
\end{array} \right) ,
E_3^s=\sqrt{1-\lambda_s}\left( 
\begin{array}{cc}
0 & 0 \\ 
\sqrt{p_s} & 0 \\ 
\end{array} \right)$  \\ \hline
\end{tabular}
\caption[table1]{Kraus operators for phase damping (PD)  and generalized amplitude damping (GAD), where 
$p_s$ and $\lambda_s$ are the decoherence probabilities for the qubit $s$. }
\label{t1}
\end{table}

Both the  PD and GAD decoherence processes preserve the X form of the density operator. As a next step, we have to find out the evolved parameters $\tilde{c}_i(t)$, as defined by Eq.(\ref{nuevasc}). 
In this direction, we use Eq.~(\ref{eq:rhox}) into Eq.~(\ref{epsilon_rho}). Remarkably, the parameters $\tilde{c}_i(t)$ turn out to be independent of $\lambda_s$. Since the evolution is Markovian, we 
further take the decoherence probability $p_s = 1 - \exp(-t \,\gamma_s)$ for both PD and GAD channels. In turn, the evolution is described by the parameters displayed in 
Table~\ref{tab:table2} in terms of the decoherence time   
\begin{equation}
\tau_D=\frac{1}{\gamma_A +\gamma_B} \, . 
\end{equation}
\begin{table}[h]
\caption{\label{tab:table2} Correlation parameters $\tilde{c}_i(t)$ ($i=1,2,3$) for PD and GAD channels.}
\begin{ruledtabular}
\begin{tabular}{cccc}
 Channel & $\tilde{c}_1(t)$  & $\tilde{c}_2(t)$ & $\tilde{c}_3(t)$\\
\hline
 PD & ${c}_1exp[-t/\tau_D]$  & ${c}_2 exp[-t/\tau_D]$ & $( c_3-c_4c_5)$\\
\hline
 GAD & ${c}_1exp[-t/2\tau_D]$  & ${c}_2 exp[-t/2\tau_D]$ & $( c_3-c_4c_5) exp[-t/\tau_D]$\\
\end{tabular}
\end{ruledtabular}
\end{table}  
 \begin{figure}[ht!]
 \includegraphics[scale=0.34]{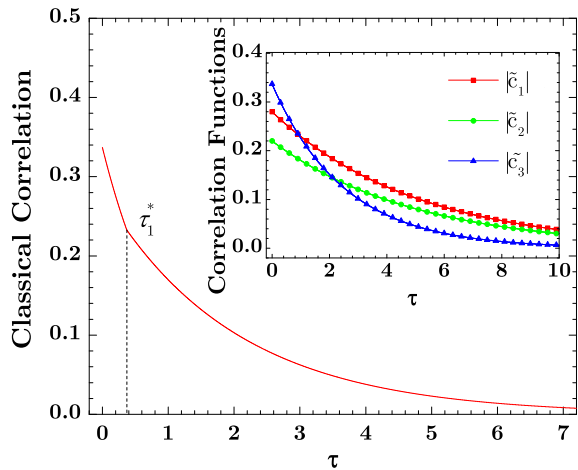} 
  \vspace{-0.8cm}
\caption{(Color online) Classical correlation as a function of $\tau=(\gamma_A+\gamma_B)t$ for a two-qubit system under the GAD channel. The initial state is in the X form, where the values for $c_i$ are selected to 
show the behavior of the sudden transition, with $c_1=0.28$, $c_2=0.22$, $c_3=0.40$, $c_4=0.10$, and $c_5=0.60$. A sudden transition in $C_G$ 
occurs at $\tau^{*}_1=0.37$. In the inset, we show the correlation parameters $|\tilde {c}_1|$, $|\tilde {c}_2|$, and $|\tilde {c}_3|$.}
  \label{fig:GAD}
\end{figure}

Then, we can directly obtain the dynamics of classical correlations $C_G(\rho_X(t))$, as given by Eq.~(\ref{cg}).  
It can be observed from Table~\ref{tab:table2} that both $\vert{c}_1(t)\vert$ and $\vert{c}_2(t)\vert$ 
display the same decay rate, which means that they do not cross as functions of time. Therefore only the crossings allowed are for 
$\vert{c}_1(t)\vert=\vert{c}_3(t)\vert$ and $\vert{c}_2(t)\vert=\vert{c}_3(t)\vert$, implying at most a single nonanalyticity (sudden change) 
in the geometric classical correlation. This 
conclusion holds for both PD and GAD channels. Indeed, a necessary and sufficient condition for sudden change in the case of PD and GAD channels are  
$\tilde{{c}}_- = \vert \tilde{{c}}_3 \vert \neq 0$ and $\tilde{{c}}_+ = \vert \tilde{{c}}_3 \vert \neq 0$, respectively.   
Therefore, the generalization of the initial state to an X state does not allow for further 
sudden changes in the classical correlation. This sustains the result that double sudden changes is an exclusive feature of quantum correlations, 
as discussed for Bell-diagonal states in Ref.~\cite{PRL111,PRL87}. We illustrate this behavior in Fig.~~\ref{fig:GAD}, where we plot $C_G$ as a function of the 
 dimensionless time $\tau = (\gamma_A+\gamma_B)\,t$ for a mixed X state under the GAD channel.  It can be observed that a single sudden transition occurs at 
$\tau^{*}_{1}=0.37$, which can be determined from the correlation parameters $c_i(t)$ in Table~\ref{tab:table2}. 

\subsection{Pointer basis for Markovian dynamics}

 Let us now apply the classical correlation $C_G$ for X states to investigate the emergence of the pointer basis of a quantum apparatus ${\cal A}$ 
subject to decoherence in a Markovian regime. The apparatus ${\cal A}$ measuring a system ${\cal S}$ suffers decoherence through the 
contact with the environment, which implies in its relaxation to a possible set of classical states known as the pointer basis~\cite{zurek}. 
As a consequence,  the information about ${\cal S}$ turns out to be accessible to a classical observer through the pointer basis associated with the apparatus. 
The emergence of the pointer basis occurs for an instant of time $\tau_E$ at which the classical correlation between ${\cal A}$ and ${\cal S}$ becomes constant~\cite{Prl109,PRL87,PRL111}. 
Therefore, we will consider a composite system $\cal{AS}$ under decoherence described by the density operator given by  Eq. (\ref{eq:rhox}). The classical correlation can be used to 
characterize the time $\tau_E$ when the pointer states  emerges, which exactly corresponds to the instant of time at which $C_G (t)$ shows a sudden transition to a constant function.

\begin{figure}[h]
 \centering
 \includegraphics[scale=0.34]{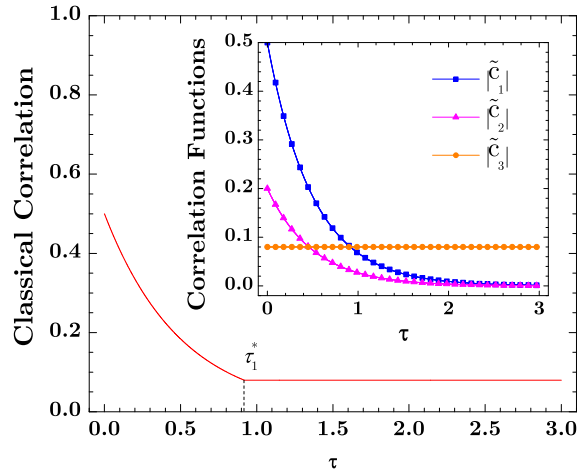} 
  \vspace{-1cm}
\caption{(Color online) Classical correlation as a function of  $\tau=(\gamma_A+\gamma_B)t$ for a two-qubit system under the PD channel. The initial state is in the X form, where 
$c_1=0.50$, $c_2=0.20$, $c_3=0.10$, $c_4=0.10$, and $c_5=0.20$, with these values chosen to illustrate the emergence of the pointer basis. This occurs at $\tau_1^{*} = 0.92$, i.e., 
$\tau_E = 0.92 \, \tau_D$. In the inset, we detail the evolution of the correlation parameters $|\tilde {c}_1|$, $|\tilde {c}_2|$, and $|\tilde {c}_3|$.}
\label{fig:PD}
\end{figure}

For the GAD channel, there is no emergence of pointer basis at a finite time, since no 
decay of $C_G$ to a constant function of time is possible. On the other hand, for the PD channel, we can analytically determine $\tau_E$. Indeed, from 
Table~\ref{tab:table2}, $C_G(t)$ gets constant after a sudden transition at finite time given by 
 \begin{equation}\label{tbasepunt}
\tau_E =  \tau_D \ln \left[ \dfrac{\tilde{c}_{+}}{ |\tilde{c_{3}}|} \right].
 \end{equation}
Comparing $\tau_E$ with the decoherence time scale $\tau_{D}$, we can observe that the pointer basis may emerge at a time smaller or larger than $\tau_{D}$. This generalizes 
the result obtained in Refs.~\cite{Prl109,PRL87,PRL111} for Bell-diagonal states. To illustrate the emergence of the pointer basis, we plot in Fig.~\ref{fig:PD} 
the decay of the classical correlation as a function of $\tau = (\gamma_A+\gamma_B)\,t$ under the PD channel for an initial state in the X form. 
The emergence of the pointer basis through the behavior of $C_G$ occurs 
then at $\tau^{*}_1=0.92$, i.e., $\tau_E = 0.92 \, \tau_D$.

\subsection{Non-Markovian dynamics}
We now consider the classical correlations for X-states in a non-Markovian open quantum system under the PD channel.  Non-Markovian dynamics describes many physical situations, e.g. 
single flourescent systems hosted in complex environments, superconducting qubits, dephasing in atomic and molecular physics, among others~\cite {Budini:7273,Falci:94,Wong:63}. 
For this work the non-Markovianity of the evolution will be handled in the local time framework developed in Ref.~\cite{AAPRA74}.  In this scenario, we start by supposing a quantum process 
governed by a Markovian master equation 
\begin{equation}
\displaystyle\frac{d \rho}{dt} = {\cal L} [\rho(t)], 
\end{equation}
where the generator ${\cal L}$ is given by
\begin{equation}
 {\cal L}[\bullet ]   = -i\left[H,\bullet \right]  + \sum_i \gamma_i \left( A_i \bullet A^\dagger_i - \frac{1}{2} \left\{A^\dagger_i A_i,\bullet \right\}\right),
 \end{equation}
with $H$ denoting the effective system Hamiltonian, $A_i$ the Lindblad operators, and $\gamma_i \ge 0$  the relaxation rates~\cite{Lindblad}. 
In order to generalize the treatment to the non-Markovian regime, the density matrix $\rho_s(t)$ of the system is written as
\begin{equation}
\rho_S(t) = \sum^{R_{max}}_{R=1}\rho_R(t), 
\end{equation}
where each auxiliary (unnormalized) operator $\rho_R$ defines the system dynamics given that the reservoir is in the R-configurational bath state, with $R_{max} $ the number of 
configurational states of the environment. The probability $P_R (t)$ that the environment is in a given state at time $t$ reads
\begin{equation}\label{Eq:prob}
P_R(t)  =  \mathrm{tr}[\rho_R(t)].
\end{equation}
We note that the set of states $\{\rho_R(t)\}$ encodes both the system dynamics and the fluctuations of the environment~\cite{AAPRA74,HPPRA75}. When the transitions 
between the configurational states do not depend on the system state, the fluctuations between the configurational states are governed by a classical master 
equation~\cite{VKampen}, with a structure following from Eq.~(\ref{Eq:prob}). This kind of environmental fluctuations are called self-fluctuating environments.   
For our work, we restrict our attention to a two-qubit system $A$ and $B$ interacting with a self-fluctuating environment. Then, we model the environment as 
being characterized by a two-dimensional configurational space ($\mathrm{R}_{max}=2$), which only affects the decay rates of the system.  Each state follows by itself a Markovian 
master equation
\begin{eqnarray}\label{Eq:rho1}
\nonumber \frac{d\rho_1(t)}{dt}&=& -i[H_1, \rho_1(t)] + \gamma^{A}_1(\mathcal {L}^{A}[\rho_1(t)]) + \gamma^{B}_1(\mathcal {L}^{B}[\rho_1(t)]) \\&-& \phi_{21}\rho_1(t)+\phi_{12}\rho_{2}(t),
\end{eqnarray}
\begin{eqnarray}\label{Eq:rho2}
\nonumber \frac{d\rho_2(t)}{dt} &=& -i[H_2, \rho_2(t)] + \gamma^{A}_2(\mathcal {L}^{A}[\rho_2(t)]) + \gamma^{B}_2(\mathcal {L}^{B}[\rho_2(t)]) \nonumber\\ &-& \phi_{21}\rho_2(t)+\phi_{12}\rho_{1}(t),
\end{eqnarray}
where the structure of the superoperator $\mathcal {L}$ for the PD channel is given by 
\begin{equation}\label{Eq:NMPD}
\mathcal {L}^{A,B}[\bullet ]=(\sigma^{A,B}_z\bullet\sigma^{A,B}_z - \bullet).
\end{equation}
The first line of Eqs.~(\ref{Eq:rho1}) and (\ref{Eq:rho2}) defines  the unitary and dissipative dynamics for the two-qubit system, given that the bath is in the configurational state 
1 or 2, respectively.  The constants $\{\gamma^{A}_{1,2},\gamma^{B}_{1,2}\}$ are the natural decay rates of the system associated with each reservoir state ~\cite{nielsen}. 
The positivity of the density matrix will be ensured as long as these decoherence coefficients obey $\gamma^{A,B}_i \geq 0$~\cite{AAPRA74,budini43}.
 
On the other hand, the second line of Eqs. (\ref{Eq:rho1}) and (\ref{Eq:rho2}) describes transitions between the configurational states  of the environment (with rates $\phi_{12}$ 
and $\phi_{21}$)~\cite{budini43}. For a matter of simplicity, the decay rates associated with each subsystem will be chosen to be the same, namely, 
$\gamma^{A}_{1}= \gamma^{B}_{1}\equiv \gamma_{1}$ and $\gamma^{A}_{2}= \gamma^{B}_{2}\equiv \gamma_{2}$. Moreover, we define the characteristic dimensionless parameters  
\begin{equation}
\epsilon = \frac{\gamma_1}{\gamma_1 + \gamma_2},\quad \epsilon \in [0,1], \label{epsilon}
\end{equation}
\begin{equation}
\eta = \frac{\phi_{12}}{\phi_{12}+ \phi_{21}},\quad \eta \in [0,1], 
\end{equation}
\begin{equation}
v = \frac{ \phi_{12}+ \phi_{21}}{\gamma_1 + \gamma_2},\quad v \in [0,\infty). \label{velocity}
\end{equation}

\begin{figure}[h]
 \centering
 \includegraphics[scale=0.34]{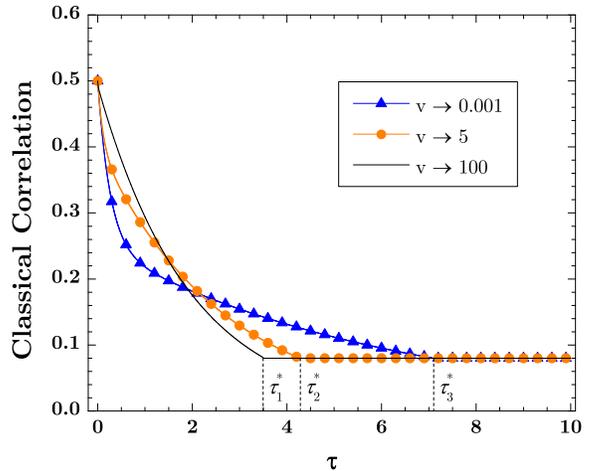}
 \vspace{-1cm}
\caption{(Color online) Classical correlation as a function of $\tau= (\gamma_1+\gamma_2)t$ for a two-qubit system under the non-Markovian PD channel.  The initial state is in the X form, 
with  $c_1= 0.50$, $c_2=0.20$, $c_3=0.10$, $c_4=0.10$, and $c_5=0.20$. We have also taken $\epsilon=0.92$ and $\eta=0.10$. The emergence times $\tau_E$ are associated 
with $\tau^{*}_1=3.5$, $\tau^{*}_2=4.3$,  and $\tau^{*}_3=7.1$. The initial state has been chosen to show the emergence of the pointer basis in the non-Markovian regime.}
\label{fig:nMPDv}
\end{figure}

Then, we characterize the evolutions given by Eqs.~(\ref{Eq:rho1}) and (\ref{Eq:rho2}) by observing that the non-Markovian PD process preserves the X state form.  
Similarly as we have done in the Markovian case, we can directly obtain the dynamics of the classical correlations from Eqs.~(\ref{c1})-(\ref{c5}) and from the definition of $C_G$ 
in Eq.~(\ref{cg}). We will analyze the system in the limit of either fast or slow environmental fluctuations. The fast limit of environmental fluctuations occurs when the reservoir 
fluctuations are much faster than the average decay rates of the system, namely, $\{ \phi_{R'R}\} \gg  \{ \gamma_R \}$, which implies that the system exhibits Markovian behavior. Then, from Eq.~(\ref{velocity}), we take $v\gg 1$. 
On the other hand, when the bath fluctuations are much slower than the average decay rate, namely, $\{ \phi_{R'R}\} \ll  \{ \gamma_R \}$, the system is in the limit of slow environmental 
fluctuations. Then, from Eq.~(\ref{velocity}), we take $v \ll 1$. 
Let us now investigate the emergence of the pointer basis for the case of the non-Markovian PD channel, given by Eqs. (\ref{Eq:rho1})-(\ref{Eq:NMPD}). In this scenario, 
the classical correlation can witness the emergence time $\tau_E$, which is illustrated in Fig.~\ref{fig:nMPDv}.  Moreover, we observe that, for 
$\{ \phi_{R'R}\} \ll \{ \gamma_R \}$, the classical correlation displays a bi-exponential decay. On the other hand, for $\{ \phi_{R'R}\} \gg  \{ \gamma_R \}$, 
the classical correlation shows a single exponential decay, such as expected for a Markovian behavior.  In addition, we can observe the emergence of the pointer 
basis for any $v$ through the sudden transitions, with $\tau_E$ greater for slower environmental fluctuations. 

\begin{figure}[h]
 \centering
 \includegraphics[scale=0.34]{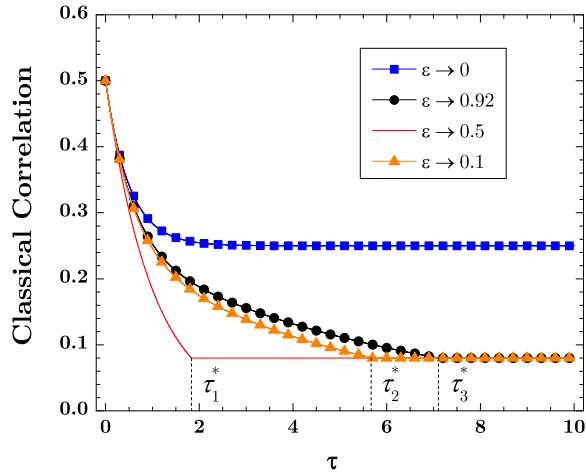}
 \vspace{-1cm}
\caption{(Color online) Classical correlation as a function of $\tau= (\gamma_1+\gamma_2)t$ for for a two-qubit system under the non-Markovian PD channel in the limit of slow fluctuations, 
with $v=0.001$ and $\eta=0.70$.  The initial state is in the X form, with $c_1= 0.50$, $c_2=0.20$, $c_3=0.10$, $c_4=0.10$,and $c_5=0.20$. The emergence times $\tau_E$ are associated with 
$\tau^{*}_1=1.8$ , $\tau^{*}_2=5.7$,  $\tau^{*}_3=7.1$.}
\label{fig:nMPD}
\end{figure}
By focusing attention on the slow configurational transitions, we show in Fig.~\ref{fig:nMPD} that $\tau_E$ strongly depends on $\epsilon$, i.e., on the ratio of decay 
rates $\gamma_1$ and $\gamma_2$. The shortest emergence time occurs for the central value $\epsilon=0.5$, where decay rates obey $\gamma_1 =\gamma_2$. 
As we move away from $\epsilon = 0.5$,  the emergence of the pointer basis is delayed. In particular, for the limit cases $\epsilon=0$ or $\epsilon=1$, the system shows a soft decay, with no sudden transition 
at finite time.

\section{conclusions}
In summary, we have analytically evaluated the trace-distance classical correlations for the case of two-qubit systems described by X states. 
In addition, we have shown the applicability of such correlations to investigate the dynamics of open quantum systems through the characterization of the pointer 
basis of an apparatus suffering either Markovian or non-Markovian decoherence. Since the non-Markovianity brings a flow of information from the environment 
back to the system during its evolution, the pointer basis has been found to emerge in a delayed time in comparison with the Markovian behavior. 
The experimental characterization of such delay in the emergence time can be achieved by a similar approach as used in Refs.~\cite{PRL111,Prl109} for 
Markovian evolutions. 
It is also remarkable to observe that, differently from the case of Bell-diagonal states, sudden transitions of entropic correlations for X 
states have been conjectured to display zero measure~\cite{Pinto:13}, which may compromise a precise 
characterization of the pointer basis. Our geometric approach avoids this obstacle, since actual sudden changes are shown to be typical for general X states. 
This may have further implications in the characterization of quantum phase transitions through geometric classical correlations.

\begin{acknowledgments}
This work is supported by the Brazilian agencies 
CNPq, CAPES, FAPERJ, and the Brazilian National Institute for Science and Technology of Quantum Information (INCT-IQ).

\end{acknowledgments}


\end{document}